\documentclass[12pt,epsf]{article}

\usepackage{graphicx,graphics,cite}
\usepackage{subfigure}
\newcommand{\ba}{\begin{array}}
\newcommand{\ea}{\end{array}}

\newcommand{\nn}{\nonumber}
\newcommand{\be}{\begin{equation}}
\newcommand{\ee}{\end{equation}}
\newcommand{\bea}{\begin{eqnarray}}
\newcommand{\eea}{\end{eqnarray}}

\def\als{\alpha_{\rm s}}
\def\siml{{\ \lower-1.2pt\vbox{\hbox{\rlap{$<$}\lower6pt\vbox{\hbox{$\sim$}}}}\ }} 

\newcommand{\MS}{\overline{\rm MS}}
\newcommand{\RS}{\rm RS}

\begin{document}
\title{\vskip-3cm{\baselineskip14pt
\centerline{\normalsize UB-ECM-PF-05-17 \hfill}
\centerline{\normalsize  August 2005\hfill}
}
\vskip1.5cm
Fit to the Bjorken, Ellis-Jaffe and Gross-Llewellyn-Smith sum rules in a
renormalon based approach}
\author{Francisco Campanario$^{1}$ and 
Antonio Pineda$^{2}$\\[0.5cm]
{\small ${}^1$ \it Department of Physics, University of Toronto,} \\
{\small \it St. George Street, Toronto, Canada M5S 1A7} \vspace{0.3cm}\\
{\small ${}^2$ \it Dept. d'Estructura i Constituents de la Mat\`eria, U. Barcelona,}\\
{\small\it Diagonal 647, E-08028 Barcelona, Catalonia,  Spain}}
\date{}

\maketitle

\thispagestyle{empty}

\begin{abstract}
We study the large order behaviour in perturbation theory of the 
Bjorken, Ellis-Jaffe and Gross-Llewellyn-Smith sum rules. In particular, 
we consider their first infrared renormalons, for which we obtain 
their analytic structure with logarithmic accuracy and also an 
approximate determination of their normalization constant. 
Estimates of higher order terms of the perturbative series are 
given. The Renormalon subtracted scheme is worked out for these 
observables and compared with experimental data. Overall, 
good agreement with experiment is found. This allows us to 
obtain ${\hat a}_0$ and some higher-twist non-perturbative constants from 
experiment: ${\hat a}_0=0.141\pm 0.089$; 
$f_{3,\RS}(1\;{\rm GeV})=-0.124^{+0.137}_{-0.142}$ GeV$^{2}$.
\\[2mm]
\end{abstract}

\newpage

\section{Introduction}
\label{int}

Deep Inelastic Scattering (DIS) is one of the few places where one 
can test, on a very solid theoretical ground, asymptotic freedom 
and the operator product expansion against experiment. This is so because,
for the moments of the different structure functions, the 
transfer momentum lies in the euclidean region and far away of the 
physical cuts (for $Q^2$ large). Therefore, the theoretical predictions do not rely on 
any kind of quark-hadron duality and may provide with solid and 
very clean determinations of $\als$ and some non-perturbative matrix elements. 
Equally interesting is the study of the interplay between the perturbative and 
non-perturbative regime. This can only be done if a full control on the perturbative 
series is achieved.

Within DIS, special consideration deserve the sum rules for which the 
matrix elements are related to symmetry generators (in this case they 
can be computed in absolute value within perturbation theory), 
like the Gross-Llewellyn-Smith (GLS) sum rule~\cite{GLS69},  
or to some low energy constants (which can be directly measured from 
experiment), like the Bjorken sum rule~\cite{Bjorken}. In this paper, 
we will concentrate on the Bjorken, Ellis-Jaffe~\cite{EllJaf74} and 
GLS sum rules. Their leading-twist term has 
been computed with next-to-next-to-leading order (NNLO) accuracy and they  
have been measured with increasingly good accuracy over the years. 

By the Operator Product Expansion, the short- and long-distance contributions are separated 
and a sum rule, M, can be expressed in the following way:
\be
\label{sum1}
M= C{<}J{>}+ B\frac{{<}R{>}}{Q^2}+\ldots\,,
\ee
with short-distance Wilson coefficients $C, B,$ $ \ldots$ and 
long-distance matrix elements ${<}J{>},$ ${<}R{>},$ $\ldots$.
The perturbative series of $C$ are expected to be asymptotic and, therefore, 
to diverge for a high enough order in perturbation theory. Moreover, 
in schemes without strict separation of large and small momenta,
such as $\overline{\mathrm{MS}}$, they are believed to be non-Borel 
summable.  This is because 
when calculating the matching coefficients $C$, \dots,
the integrals run over all loop momenta, including small ones. 
Therefore, they also contain,
in addition to the main short-distance contributions,
contributions from large distances,
where perturbation theory is ill-defined.
These contributions produce infrared renormalon singularities 
\cite{tHooft},
factorially growing contributions to coefficients of the perturbative series,
which lead to ambiguities $\sim\left(\Lambda_{\mathrm{QCD}}/Q\right)^{2 n}$
in the matching coefficients $C$, \dots{}.
Similarly, matrix elements of higher-dimensional operators ${<}R{>}$, 
\dots also contain, in addition to the main large-distance contributions,
contributions from short distances,
which produce ultraviolet-renormalon singularities.
They lead to ambiguities of the order $\Lambda_{\mathrm{QCD}}^{2 n}$
times lower-dimensional matrix elements (e.g., ${<} J {>}$).
These two kinds of renormalon ambiguities
should cancel in physical observables~\cite{Parisi:1978bj,Mueller:1993pa,
Bigi:1994em,Neubert:1995wq,Luke:1995xd,Beneke:1995rs}, in this case M.

The intrinsic (minimal) error associated to the perturbative series is of the 
order of the higher twist correction. Thus, one can not unambiguously 
determine the higher twist terms, unless a prescription to deal with the 
perturbative series that has power-like accuracy is given. In this paper 
we will adapt to this case the prescription used for 
heavy quark physics in Refs. \cite{Pineda:2001zq,Pineda:2002se,Bali:2003jq}. 
The idea is that the leading divergent behaviour of the perturbative series is related 
to the closest singularities 
in the Borel plane of its Borel transform. In heavy quark physics, they lie on the positive axis 
(infrared renormalons). In the case of the sum rules considered in this paper, the closest singularities 
lie on the positive and negative axis at equal distance to the origin. We will assume that the one in the 
positive axis will dominate the asymptotics of the perturbative series. 
Since these singularities cancel against the ultraviolet renormalons of the low energy dynamics of the twist-4 
operators, the proposal will be to shift the singularities from the perturbative series to the twist-4 operators. 
We will refer to this prescription as the Renormalon Subtracted (RS) scheme and apply it to the 
Bjorken, Ellis-Jaffe and GLS sum rules. 
We will obtain the contribution of the leading infrared renormalon, subtract it from
the perturbative series and add it to the low energy matrix elements, in our case, to the twist-4 
operators. In this way one enlarges the range of convergence of 
the Borel transform of the perturbative series, which can be defined with power-like 
accuracy. This procedure has proven to be extremely successful in heavy quark physics, 
where it has been shown that perturbation theory works very well and good determinations 
of non-perturbative subleading corrections have been obtained using either 
lattice or experimental data.
One should be aware, however, that in heavy quark physics one was in an optimal 
situation, since the singularity in the Borel plane was quite 
close to the origin ($u=1/2$). We are now going to be in a less optimal situation, 
since the closest singularities lie at $u=\pm 1$. The physical 
situation is also completely different since now we are talking 
of a system made with light fermions. Therefore, it is interesting to investigate 
if a similar improvement is obtained in this case. We will do so in this paper.

The paper is organized as follows. In the next section we will introduce the relevant sum rules.
In Sect. 3 the Borel transform of the first infrared renormalon of the leading-twist Wilson coefficient 
will be calculated with leading log accuracy, as well as the normalization constant and estimates of the 
higher order terms of the perturbative series.  
The RS scheme will be worked out in Sect. 4.
In Sect. 5 the comparison with the experimental data will be done allowing us the extraction of 
some non-perturbative matrix elements. Finally, the conclusions are presented in Sect. 6.
\section{Sum Rules}
The Bjorken and Ellis-Jaffe sum rules are related to polarized deep inelastic electron-nucleon 
scattering, which is described by the hadronic tensor
\begin{eqnarray}
 W_{\mu\nu} & = & \frac{1}{4\pi} \int d^4z e^{iqz} \langle p,s|
 J_\mu(z) J_{\nu}(0) |p,s\rangle \nonumber \\
 & = & \left( -g_{\mu\nu} +\frac{q_\mu q_\nu}{q^2} \right) F_1(x,Q^2)
   + \left( p_\mu-\frac{p\cdot q}{q^2} q_\mu \right) 
       \left( p_\nu - \frac{p\cdot q}{q^2} q_\nu \right) 
      \frac{1}{p\cdot q} F_2(x,Q^2)  \nonumber \\
 & & + i \epsilon_{\mu\nu\rho\sigma} q_{\rho} \left(
        \frac{ s_\sigma}{p\cdot q} g_1(x,Q^2)
         +\frac{s_\sigma p\cdot q - p_\sigma q\cdot s}{ (p\cdot q)^2}
         g_2(x,Q^2) \right)
\,.
\end{eqnarray}
Here $J_\mu = \sum_{i=1}^{n_f} e_i \overline{\psi_i} \gamma_\mu \psi_i$ is the 
electromagnetic quark current where $ e_i =2/3,-1/3,-1/3,\cdots $ 
is the electromagnetic charge of a quark with the corresponding flavour, 
$u$, $d$, $s$.
$x=Q^2/(2p\cdot q)$ is the Bjorken scaling variable and $Q^2 = -q^2$ is the
square of the transferred momentum. $ |p,s\rangle $ is the
nucleon state that is normalized as $\langle p,s | p', s' \rangle$ 
$= 2 p^0 (2\pi)^3 \delta^{(3)}(p-p') \delta_{ss'}$. The polarization vector
of the nucleon is expressed as 
$s_\sigma = \overline{U}(p,s) \gamma_\sigma \gamma_5 U(p,s)$ where $U(p,s)$
is the nucleon spinor $\overline{U}(p,s) U(p,s) = 2 m_N$.

The Ellis-Jaffe sum rule then reads
\bea
\label{EJ}
&&\hspace{-0.8cm}M_1^{p/n}(Q^2)
=\left(\pm\frac{1}{12} g_A+\frac{1}{36}a_8\right)C_{B}(\als)
+\frac{1}{9}{\hat a_0}
C_{EJS}(\als)
\\     
\nonumber
&&\hspace{-0.8cm}-\frac{8}{9Q^2}\left[\left(\pm\frac{1}{12} f_3(Q_0)+
\frac{1}{36}f_8(Q_0)\right)
\left(\frac{\alpha_s(Q_0^2)}{\alpha_s(Q^2)}\right)^
{-\frac{\gamma_{NS}^0}{2\beta_0}}\hspace{-0.1cm}\left(1+{\cal O}(\als)\right)+
\right.  \\
\left. \right.
&& 
\left.
+\frac{1}{9}f_0(Q_0)
\left(\frac{\alpha_s(Q_0^2)}{\alpha_s(Q^2)}\right)^
{-\frac{1}{2\beta_0}
(\gamma_{NS}^0+\frac{4}{3}N_f)}\left(1+{\cal O}(\als)\right)
\right]
+{\cal O}\left({1 \over Q^4}\right),\nn 
\eea
and the Bjorken sum rule is the difference between the proton and 
neutron sum rule:
\be
\label{Bjorken}
M_1^B\equiv M_1^p(Q^2)-M_1^n(Q^2)={g_A \over 6}C_B(\als)
-{4 \over 27}{1 \over Q^2}f_3(Q_0)
\left[{\als(Q_0^2)\over \als(Q^2)}\right]^{-\gamma_{\rm NS}^0 \over 2\beta_0}
\left(1+{\cal O}(\als)\right)
+{\cal O}\left({1 \over Q^4}\right)
\,,
\ee
where the definitions are the following:
\be
\gamma_{NS}^0={16 \over 3}C_F; \qquad 
\gamma_{S}^0=\gamma_{NS}^0+{4 \over 3}n_f; 
\ee
\be
C_X(Q)=1+\sum_{s=0}^{\infty}C_X^{(s)}\als^{s+1}(\nu)\,,
\label{pertubativeseries}
\ee
and  $X
=\{B$, $EJ$, $GLS\}$, the latter to be defined below.
For the Bjorken sum rule, the $\alpha_s$ correction~\cite{bj1loop},
the $\alpha_s^2$ correction~\cite{bj2loop},
and the $\alpha_s^3$ correction~\cite{bj3loop}
have been calculated in the leading twist approximation. 
Higher twist corrections have also been calculated~\cite{highertwists}. 
The Ellis-Jaffe sum rule 
for the proton and neutron was calculated to order
$\alpha_s$~\cite{kod}, to order $\alpha_s^2$
\cite{ej2loop} and to order $\alpha_s^3$~\cite{Larin:1997qq} 
in the leading twist approximation. 
Power corrections were calculated in~\cite{powercorrections}.
The LO renormalization group running of the twist-4 operators have been computed 
in Ref.~\cite{Kawamura:1996gg}. The non-perturbative matrix elements are defined 
in the following way:
\begin{equation}
\label{gaa8}
 \begin{array}{llrll}
 |g_A| s_\sigma & = & 2 \langle p,s | J^{5,3}_\sigma | p,s \rangle & = &
         (\Delta u - \Delta d ) s_\sigma , \nonumber \\
 a_8 s_\sigma & = & 2 \sqrt{3} \langle p,s | J^{5,8}_\sigma | p,s \rangle
 & = & ( \Delta u + \Delta d - 2 \Delta s) s_\sigma , \nonumber \\
 a_0(\mu^2)s_\sigma & = & \langle p,s | J^{5}_\sigma | p,s \rangle
 & = & ( \Delta u + \Delta d +  \Delta s) s_\sigma.
\end{array} 
\end{equation}
where $J_\sigma^{5,a}(x) = \overline{\psi} \gamma_\sigma
 \gamma_5 t^{a} \psi(x)$ is the non-singlet axial current, where $t^a$ is a
generator of the flavour group, and $J_\sigma^5(x) =
  \sum_{i=1}^{n_f} \overline{\psi_i} \gamma_\sigma \gamma_5 \psi_i(x)$
is the singlet axial current. $|g_A|$ is the absolute value of the constant of 
the neutron beta-decay, $|g_A/g_V| = F+D=1.2695 \pm 0.0029$ 
\cite{Eidelman:2004wy}.
$a_8=3F-D=0.572 \pm 0.019$ is the hyperon decay constant\footnote{We will 
obtain this number from Hyperon decays, see~\cite{Ratcliffe:2004jt}. $F/D=2/3$ in the large $N_c$.}. The matrix element of the singlet 
axial current $a_0(\mu^2)$ will be redefined in a proper
invariant way as a constant ${\hat a}_0$:
\be
{\hat a}_0=
{\rm exp}
\left(
-2\int^{\als(\nu)}d\als^{\prime}{\gamma^s(\als^{\prime}) \over \beta(\als^{\prime})}
\right)
a_0(\nu)
\,.
\ee
We use the  notation
$ \Delta q(\mu^2) s_\sigma = \langle p,s | \overline{q} \gamma_{\sigma}
 \gamma_5 q |p,s \rangle$, $q = u,d,s$,
for the polarized quark distributions.
$f_0$, $f_3$ and $f_8$ are the twist-4 counter parts of $a_0$, $a_3$ 
and $a_8$. $f_i$'s are scale dependent and here they are defined at $Q_0^2$, 
i.e. $f_i$ is the reduced matrix element of $R_{2\sigma}^i$,
renormalized at $Q_0^2$, which is defined for the general flavor
indices, with $t^i$ being the flavor matrices, as 
\be
R_{2\sigma}^i=g\overline{\psi}\tilde{G}_{\sigma\nu}\gamma^{\nu}t^i\psi,
\quad \langle p,s|R_{2\sigma}^i|p,s \rangle=f_i s_{\sigma} 
\quad (i=0,\cdots,8),
\ee
and $\tilde{G}_{\mu\nu}=\frac{1}{2}\varepsilon_{\mu\nu\alpha\beta}G^{\alpha\beta}$
is the dual field strength.

In the left-hand side of equations~(\ref{EJ}-\ref{Bjorken}) target-mass effects have been included using the Nachtmann variable~\cite{Nachtmann:1973mr}. They read
\bea
    M_1^N(Q^2)
    &\equiv& \int_0^1 dx {\xi^{2} \over x^2} 
	\left\{ g_1^N(x, Q^2) \left[ {x \over \xi} - 
    {1 \over 9} {m_N^2 x^2 \over Q^2} {\xi \over x}
    \right] \right. 
	-\left. g_2^N(x, Q^2) ~ {m_N^2 x^2 \over Q^2} {4 \over 3} \right\} ~~~~
	\nonumber\\
    && 
	= \int_0^1 g_1^{N}(x,Q^2)dx + {\mu_4^N \over Q^2}+{\cal O}\left({1 \over Q^4}\right)
	= \Gamma_1^N(Q^2) + {\mu_4^N \over Q^2}+{\cal O}\left({1 \over Q^4}\right)
    \label{eq:i_nm1}
 \eea
where $\xi = 2x / \left( 1 + \sqrt{1 + 4 m_N^2 x^2 / Q^2} \right)$ is the
Nachtmann scaling variable, $m_N$ is the nucleon mass. The quantity $M_1$
is the first Nachtmann moment of $g_1$ that absorbs all the
target mass corrections, $\sim (m_N^2/Q^2)^n$, and 
\begin{eqnarray}
\mu_4^{N}
&=& -\frac{m_N^2}{9}
    \left( a^{N}_2 + 4 d^{N}_2  \right),
\end {eqnarray}
where $a_2^{N}$ is the target mass correction given by the
$x^2$-weighted moment of the leading-twist $g_1$ structure function,
and $d_2^{N}$ is a twist-3 matrix element given by
\begin{eqnarray}
d^{N}_2
&=& \int_0^1 dx~x^2 \left( 2g^{N}_1 + 3g^{N}_2 \right).
\end{eqnarray}
We will consider the Ellis-Jaffe proton sum rule and the Bjorken sum rule. 
For the former we will use the data points given in Ref.~\cite{Osipenko:2005nx}, which are already 
given in terms of $M_1^p$, and for the latter we will use the data points given in 
Ref.~\cite{Deur:2004ti}, which used the values\footnote{To take them as constants is an approximation, nevertheless, 
their effect on the fit is small in comparison with other source of errors.}:
\be
d_2^{p-n}=-0.0029\,, \qquad a_2^{p-n}=0.0279\,.
\ee

\medskip

If one also considers DIS of neutrinos with nucleons, 
the GLS sum rule appears, for which the 
leading twist have been computed with NNLO accuracy (see for instance 
\cite{Hinchliffe:1996hc}):
\be
\label{GLS}
M_3^{GLS} \equiv
{1 \over 2}
\left(
M_3^{p}(Q^2)
+
M_3^{n}(Q^2)
\right)=3C_{GLS}(\als)
-\frac{8}{9}
\left(\frac{\alpha_s(Q_0^2)}{\alpha_s(Q^2)}\right)^
{-\frac{\gamma_{NS}^0}{2\beta_0}}
\frac{\langle\langle O^{{\rm S}(p+n)}_5\rangle\rangle(Q_0)}{Q^2}
\,,
\ee
where
\bea
\nn
M_3^N(Q^2) &=& \frac{1}{3} \int_0^1 dx \; F_{3}^{\nu N}(x;Q^2) 
\frac{\zeta^2}{x^2}
\left[ 1 + 2 \left( 1 + \frac{4 m_N^2 x^2}{Q^2} \right)^{1/2}
\right] 
\\
&=&  \int_0^1 dx \; F_{3}^{\nu N}(x;Q^2)
\left[ 1 - \frac{2}{3} \frac{m_N^2 x^2}{Q^2} + {\cal O}\left(
\frac{m_N^4}{Q^4} \right) \right] \ ,
\label{M3o1}
\eea
\begin{equation}
\langle P| O^{{\rm S}}_{5\mu}|P\rangle
_{{\rm \scriptstyle \hspace{-4ex}spin}
  \atop{\rm \scriptstyle averaged}}
\equiv 2P_{\mu}\langle\langle  O^{{\rm S}}_5
\rangle\rangle
\equiv 2P_{\mu}f_5^S
\,,
\end{equation}
and
\begin{equation}
O^{{\rm S}}_{5\mu} = 
\bar{u}\tilde{G}_{\mu\nu}\gamma^{\nu}\gamma_5 u
+ \bar{d}\tilde{G}_{\mu\nu}\gamma^{\nu}\gamma_5 d.
\end{equation}
For this sum rule, we will use the data of
the CCFR collaboration~\cite{Kim:1998ki}.

The 
difference between $C_{GLS}$ and $C_B$ first start at ${\cal O}(\als^3)$ and 
is proportional to the number of light fermions. This new term is 
of a "light-by-light" nature and proportional to a new casimir.  
The anomalous dimension 
of the higher twist is equal to the Bjorken case. 

The difference between the three $C_X$ is the light flavour 
dependence. In the limit $n_f \rightarrow 0$ they are all equal.

We also consider the correction due to the charm quark (with finite mass) to the 
perturbative series of $C_B$ and $C_{GLS}$. They have been computed in Ref. 
\cite{Blumlein:1998sh}. The ${\cal O}(\als^2)$ correction is equal for both of 
them and rather small (actually negligible compared with 
other source of errors). Note however that the leading order 
correction is different in each case (zero for the Bjorken case). 
This correction depends on the Cabibbo angle. We take the value $\sin \theta=0.224$~\cite{Eidelman:2004wy}. According to Ref.~\cite{Blumlein:1998sh}, 
it is a good approximation to work with 3 light flavours plus 
one massive flavour up to rather large energies. This is the situation 
we will consider in this paper.

\section{Renormalons}
The Wilson coefficients $C_X$ can be expressed in terms 
of $S_X$, their Borel transform, defined as
\begin{equation}
S_X(u)= \sum_{n=0}^\infty \frac{C^{(n)}_X}{n!}
\left(\frac{4 \pi}{\beta_0} u \right)^{n} 
\,,
\label{borel}
\end{equation}
in the following way:
\begin{equation}
C_X(Q)= 1+\frac{4 \pi}
{\beta_0}\int_0^{\infty} S_X(u) e^{\frac{-4\pi}{\beta_0 \alpha_s(\nu)}u } du
\,.
\label{laplace}
\end{equation} 
However, the perturbative expansions in $\als $ of 
the Wilson coefficients $C_X$ 
are expected to be asymptotic and non-borel sumable. In other words, we 
expect to have singularities in the real axis of $S_X(u)$. 
Those lying in the positive axis are called infrared renormalons,  
and those lying on the negative axis are called ultraviolet renormalons. 
The position and strength of the singularities can be obtained by 
using the renormalization group and consistency with the operator product 
expansion. In particular, the infrared renormalons of the 
perturbative series are obtained by demanding their cancellation 
with the ultraviolet renormalons of the higher-twist terms. 
Although the renormalon cancellation has only been explicitly shown in some cases 
in the large-$\beta_0$ limit, it is assumed to hold beyond this approximation. 
Based on this assumption, one may obtain additional information
on the structure of the infrared renormalon singularities of the matching coefficients,
based on the knowledge of the ultraviolet renormalons in higher-dimensional matrix elements,
which are controlled by the renormalization group~\cite{Parisi:1978bj}.
This model-independent approach has been applied in heavy quark effective theory 
in~\cite{Beneke:1995rs,Grozin:1997ih,Pineda:2001zq,Campanario:2003xi}.
In our case, the singularities closest to the 
origin are located in the real axis at $u=\pm 1$. 

The ultraviolet renormalon structure of the moments of the DIS structure functions have 
been computed in Ref.~\cite{Beneke:1997qd}. For those one gets the ultraviolet 
renormalon for the sum rules we are discussing here, which is the same in all 
cases up to a constant. For the case of the GLS or Bjorken sum rule, the ultraviolet renormalon 
formally dominates for $n_f > 2$ for $n \rightarrow \infty$. For the Ellis-Jaffe 
sum rule, since the infrared renormalon is weaker, the ultraviolet renormalon 
dominance appears for even a smaller number of flavours. Nevertheless, 
at low orders in perturbation theory 
the infrared renormalon appears to be dominant. This can be seen 
from the fact that the sign of the known terms of the perturbative series is 
equal whereas if the ultraviolet renormalon were to be dominant we would find a sign 
alternating series. Nevertheless, we will perform a conformal mapping to 
avoid the ultraviolet renormalon. The fact that we will obtain similar number 
than without conformal mapping will support the view that the normalization 
constant of the ultraviolet renormalon is small in comparison with the infrared one. 
Indeed, a similar conclusion was obtained in Ref.~\cite{Ellis:1995jv} using 
Pade approximants. Therefore, we will neglect ultraviolet renormalon effects in the 
leading-twist Wilson coefficients in what follows. 

The Borel transform near the closest infrared renormalon singularity has the following structure:
\begin{equation}
\label{SX}
S_X(u)= {\nu^2 \over Q^2}N_X^{(IR)}
\frac{1}{(a-u)^{1+b+b_X}}(1+d_1^X(a-u)+d_2^X(a-u)^2+\cdots) +S_{reg}(u)
\,,
\end{equation}
where $S_{reg}(u)$ is an analytic function at $u=a$ and 
we define $b={\beta_1/\beta_0^2}$. The procedure to fix the coefficients 
of this expansion (except $N_X^{(IR)}$) is by demanding consistency with the 
operator product expansion. In other words, we demand the ambiguity of the Borel 
transform   
to cancel with the ambiguity of the ultraviolet renormalons of the twist-4 matrix elements 
(see Eqs. (\ref{EJ},\ref{Bjorken},\ref{GLS})). 
Therefore,
\bea
\nn
&&{\rm Im}\left[\int_0^{\infty} S_X(u) e^{\frac{-4\pi}{\beta_0 \alpha_s(\nu)}u } du\right]
\propto
{\Lambda_{\MS}^2 \over Q^2}\als(Q)^{-b_X}
\\
&&
\qquad
\qquad
=
{\nu^2 \over Q^2}e^{\frac{-4\pi}{\beta_0 \alpha_s(\nu)}}\als(\nu)^{-b-b_X}
\left(1-{\beta_0\over 4 \pi}\als(\nu)\ln\left(\nu^2 \over Q^2\right)\right)^{b_X}
\,.
\eea
This fixes $a=1$ and $b_X$ \cite{Mueller:1993pa}:
\be
b_{GLS}=b_{B}
=-{\gamma_{NS}^0 \over 2\beta_0}\,,
\qquad b_{EJ} =-{\gamma_{S}^0 \over 2\beta_0}\,.
\ee
$b_X$ dictates the strength of the singularity. It is interesting 
to study its dependence on $n_f$. In the Bjorken and GLS sum rules, 
for $n_f \in (0,6) \Rightarrow 
1+b+b_X \in (1,2)$ 
so, formally, one could just keep the first two terms of the series in 
Eq.~(\ref{SX}), since the next term would go 
to zero for $u \rightarrow 1$. This is also the case for Ellis-Jaffe
if $n_f < 4$, otherwise one could even stick to the first term only. 

If the Wilson coefficients multiplying the 
higher twist operators were known exactly, we could also fix the 
coefficients $d_r^X$. Unfortunately, we only know their leading log running.
Nevertheless, by performing the matching at a generic scale $\nu$, we 
will be able to resum the terms of the type $(1-u)^n\ln^n(Q^2/\nu^2)$ and 
obtain the logarithmically dominant contribution to $d_r^X \sim 
\ln^r(Q^2/\nu^2)$. We obtain
\begin{equation}
S_X(u)= {\nu^2 \over Q^2}N_X^{(IR)}
\frac{1}{(1-u)^{1+b+b_X}}{}_1F_1\left(-b_X,-b-b_X,(1-u)\ln(Q^2/\nu^2)\right) 
+S_{reg}(u=1)
\,.
\end{equation}
The leading asymptotic behaviour of the perturbative series due to the 
first infrared renormalon reads
\be
\label{CX}
C_X^{(n)}
\stackrel{n\rightarrow\infty}{=} N_X\,{\nu^2 \over Q^2}\,
\left({\beta_0 \over 4\pi}\right)^n
\,{\Gamma(n+1+b+b_X) \over
\Gamma(1+b+b_X)}
\sum_{s=0}^{\infty}d_s^X{\Gamma(1+b+b_X) \over \Gamma(1+b+b_X-s)}
{\Gamma(1+b+b_X+n-s) \over \Gamma(1+b+b_X+n)}
,
\ee
where $d_0^X\equiv 1$. The logarithmically enhanced contribution to $d_s^X$ is 
known. By introducing it to Eq.~(\ref{CX}), we obtain
\be
\label{CXRG}
C_X^{(n)}
\stackrel{n\rightarrow\infty}{=} N_X\,{\nu^2 \over Q^2}\,
\left({\beta_0 \over 4\pi}\right)^n
\,{\Gamma(1+b_X+b+n) \over
\Gamma(1+b+b_X)}
{}_1F_1\left(-b_X,-b-b_X-n,\ln(Q^2/\nu^2)\right) 
.
\ee
The above expression contains subleading terms in the $1/n$ expansion. 
In the strict $1/n$ expansion, 
it simplifies to:
\be
\label{CXRG1overn}
C_X^{(n)}
\stackrel{n\rightarrow\infty}{=} N_X\,{\nu^2 \over Q^2}\,
\left({\beta_0 \over 4\pi}\right)^n
\, {n! \,n^{b_X+b} \over
\Gamma(1+b+b_X)}
\left(1+{1 \over n}\ln(Q^2/\nu^2)\right)^{b_X} 
.
\ee
This result is numerically not very different from the 
previous expression for large $n$. 
On the other hand, 
it will give more stable results than Eq.~(\ref{CXRG}) 
when working in the RS scheme for small $n$.

\medskip

With the above results we can 
identify the contribution 
to $C_X$ that comes from the renormalon. It reads
\be
\delta C_X(\nu)=\sum_{n=n^*}^{\infty}N_X\,{\nu^2 \over Q^2}\,
\left({\beta_0 \over 4\pi}\right)^n
\, {n! \,n^{b_X+b} \over
\Gamma(1+b+b_X)}
\left(1+{1 \over n}\ln(Q^2/\nu^2)\right)^{b_X}\als^{n+1}(\nu)
\,,
\ee
where $n^*$ indicates the freedom to add and subtract finite-order contributions 
in perturbation theory. 

At this stage, it is interesting to notice that $\delta C_X(\nu)$ can 
be written in the following form 
\be
\label{deltaCXRS}
\delta C_X(\nu)=\left(\als(Q) \over \als(\nu)\right)^{-b_X}
\sum_{n=n^*}^{\infty}N_X\,{\nu^2 \over Q^2}\,
\left({\beta_0 \over 4\pi}\right)^n
\, {n! \,n^{b_X+b} \over
\Gamma(1+b+b_X)}
\als^{n+1}(\nu), 
\ee
up to subleading terms. This expression will be more convenient for our 
purposes, since it has the same scale dependence 
on $Q$ as the higher twist contribution. Therefore, it 
can be moved from the leading to the subleading twist term without jeopardizing 
the $Q$ scale dependence predicted by the factorization of scales.

\subsection{Determination of the normalization constant}
\begin{figure}[h!]
\begin{tabular}{cr}
\hspace{-20pt}
\put(-10,129){$N_B$}
\put(160,118){LO}
\put(36,120){NLO}
\put(195,97){NNLO}
\put(5,1){$\nu$}
\subfigure[{$N_B$}, $N_{EJ}$ and $N_{GLS}$ with $n_f=0$.]{\includegraphics[scale
=0.8]{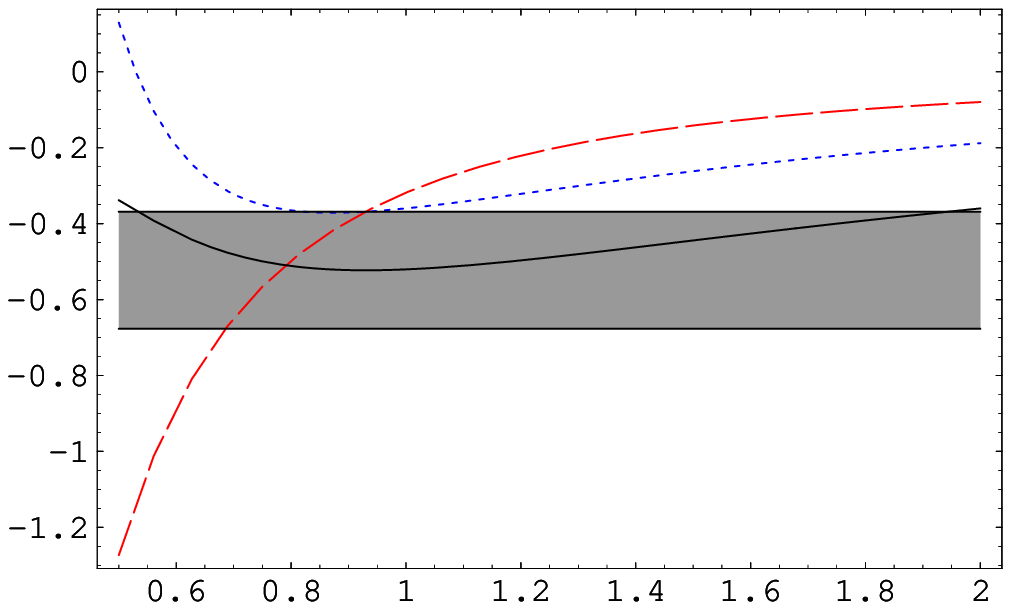}} &~
\hspace{25pt}
\put(-13,129){$N_B$}
\put(160,120){LO}
\put(160,99){NLO}
\put(190,86){NNLO}
\put(5,1){$\nu$}
\subfigure[{$N_B$} with $n_f=3$.]{\includegraphics[scale
=0.8]{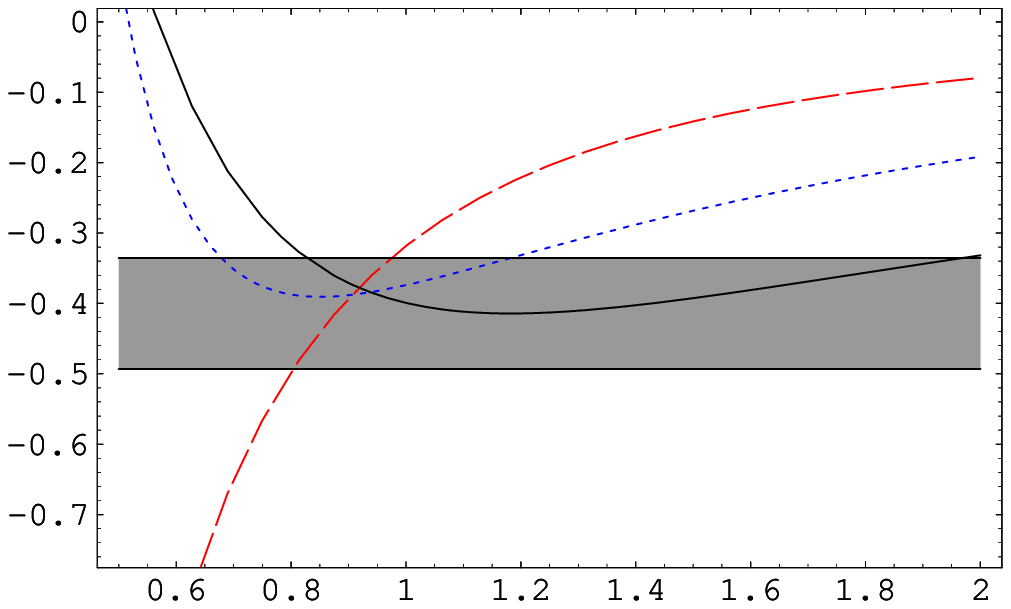}}
\\
\hspace{-20pt}
\put(-10,129){$N_B$}
\put(94,58){NLO}
\put(68,115){NNLO}
\put(46,33){LO}
\put(5,1){$\nu$}
\subfigure[{$N_B$} with $n_f=6$.]{\includegraphics[scale=0.8]{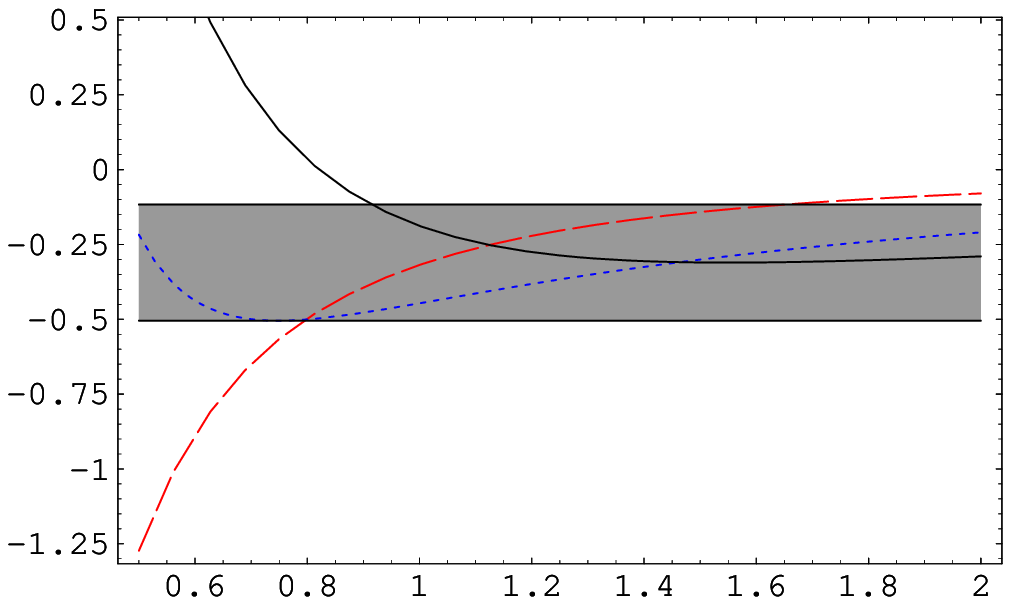}} &~
\hspace{25pt}
\put(-13,129){$N_{EJ}$}
\put(69,35){LO}
\put(60,120){NLO}
\put(40,94){NNLO}
\put(5,1){$\nu$}
\subfigure[{$N_{EJ}$} with $n_f=3$.]{\includegraphics[scale
=0.8]{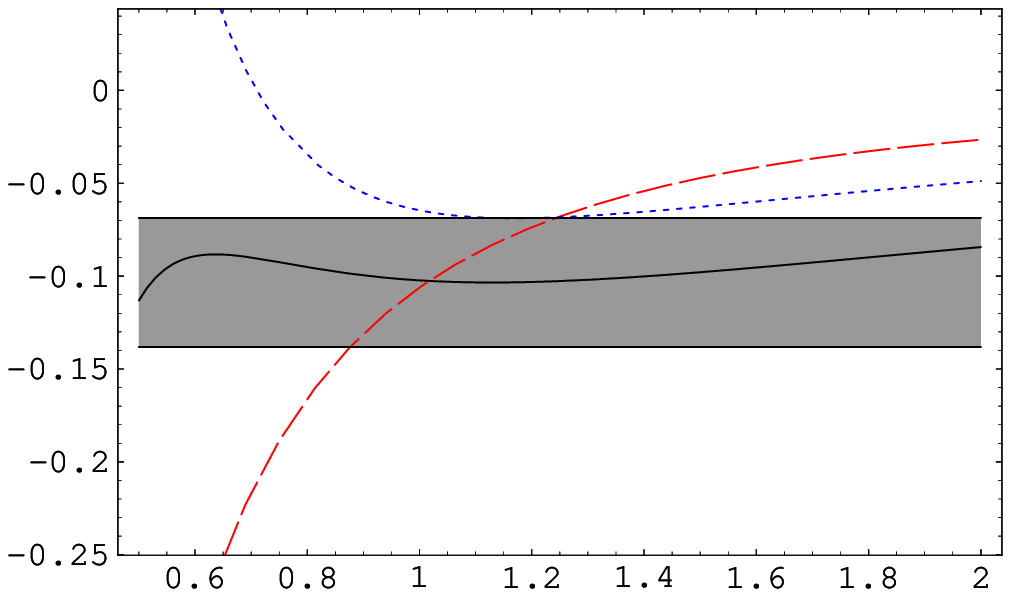}}
\\
\hspace{-20pt}
\put(-15,129){$N_{GLS}$}
\put(160,118){LO}
\put(160,96){NLO}
\put(190,83){NNLO}
\put(5,1){$\nu$}
\subfigure[{$N_{GLS}$} with $n_f=3$.]{\includegraphics[scale
=0.8]{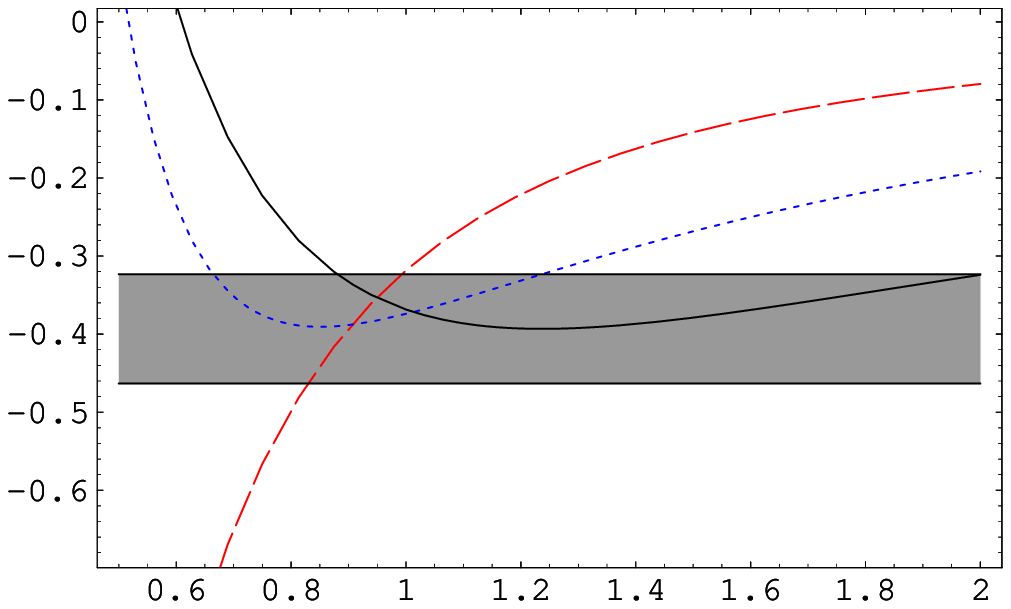}}&~
\hspace{25pt}
\put(-13,102){$N_{GLS}$}
\put(46,35){LO}
\put(87,53){NLO}
\put(67,115){NNLO}
\put(5,1){$\nu$}
\subfigure[{$N_{GLS}$} with $n_f=6$.]{\includegraphics[scale
=0.8]{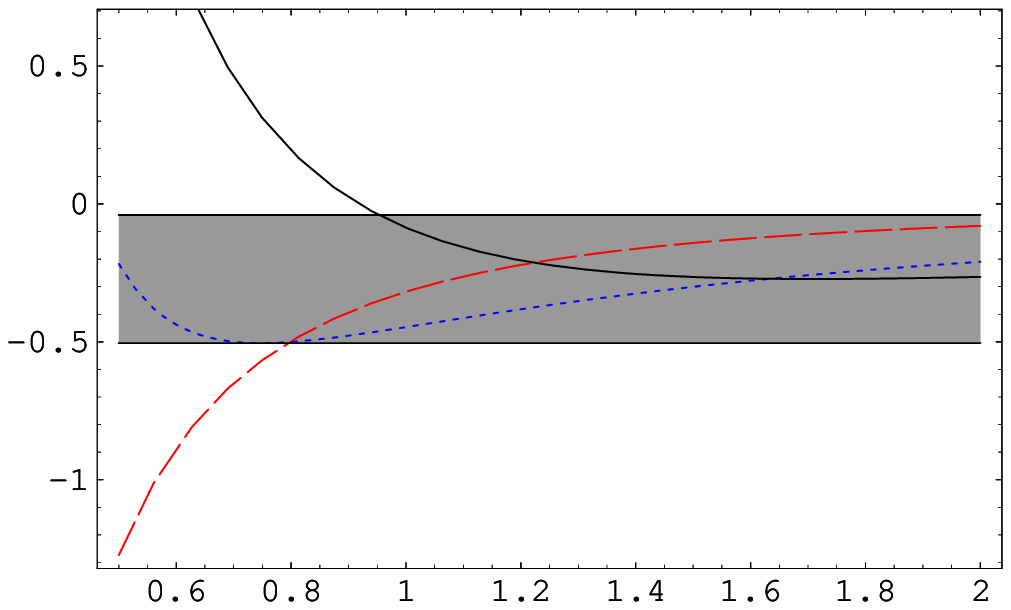}}\label{GLSnf6}
\end{tabular}
\caption {{\it Scale dependence of $N_X$ with conformal mapping for $n_f={0,3,6}$. 
The dashed (red)
line is the LO result, the dotted (blue) line is the NLO result and the 
continuous (black) line the NNLO result. The band represents the error.  
}}
\label{figNX}
\end{figure}

In this subsection we will obtain $N_X$. Let us momentarily neglect the 
ultraviolet renormalon. If this were the case, we could concentrate on the  
singularity closest to the origin in the Borel plane located at $u=1$. Then, 
we can proceed in analogy with Refs.~\cite{Lee,Pineda:2001zq,Bali:2003jq} 
and define the new function
\be
D_X(u)=(1-u)^{1+b+b_X}S_X(u)=\sum_{n=0}^{\infty}D_X^{(n)}u^n
\,.
\ee
One would then obtain $N_X$ from the following identity
\be
\label{DX}
D_X(u=1)=\sum_{n=0}^{\infty}D_X^{(n)}=N_X{\nu^2 \over Q^2}
\,,
\ee
where the first three terms of the expansion are known. Note that, formally, 
the expansion parameter is $u=1$, but one does not really know if there are small factors multiplying 
the powers of $u$. In practice, the series is quite convergent, and stable numbers for 
$N_X$ can be obtained from most of the sum rules and values of $n_f$. Nevertheless, 
we still have the problem of the ultraviolet renormalon located at $u=-1$. Formally, 
this renormalon would make the series in Eq.~(\ref{DX}) non-convergent. In order to 
avoid this problem we will perform the conformal mapping~\cite{Contreras:2002kf}, 
\be
w(u)=\frac{\sqrt{1+u}-\sqrt{1-u/2}}{\sqrt{1+u}+\sqrt{1-u/2}}
\,.
\ee
This transformation maps the first infrared renormalon to $w=1/3$ and all other singularities to 
the unit circle $|w|=1$. In the conformal mapping the expansion parameter is $w=1/3$.
In practice the effect of doing the conformal mapping is small, which points to the fact that 
the effect of ultraviolet renormalons is small in comparison with the effect of the
infrared renormalon located at $u=1$. We will only give numbers for the conformal mapping case 
for theoretical reasons.  Nevertheless, as we have already mentioned, they will be quite similar to the 
computation without conformal mapping. Our best values for $N_X$ can be found in 
Table~\ref{tableNX}. They have been computed with NNLO accuracy, after conformal mapping, 
at the scale of minimal sensitivity to the scale variation. The scale dependence of the results as well as the 
convergence is shown in  Fig.~\ref{figNX} for some selected values of $n_f$. 
We can see that in most 
cases the scale dependence becomes smother as we go to higher orders. 
The convergence depends on the number of flavours. It is optimal for $n_f=3$, 
which actually happens to be the most interesting case 
from the physical point of view and it deteriorates for large $n_f$. The error 
quoted in Table~\ref{tableNX}, and represented in Fig.~\ref{figNX} by the grey band,
stands by the maximum between the difference between the NNLO and 
NLO evaluation at the scale of minimal sensitivity of the NNLO evaluation and 
the difference between the NNLO and NLO at the scale of minimal sensitivity for 
each of them. 

The values of $N_B$ and $N_{GLS}$ are consistent with each other within errors. 
This is consistent with the interpretation that the light-by-light term does 
not contribute to the renormalon as it was done in Ref.~\cite{Contreras:2002kf}. On the other 
hand our value for $N_B$ appears to be smaller than the number obtained in 
Ref.~\cite{Ellis:1995jv}.
\begin{table}[th!]
\begin{center}
\begin{tabular}{|c|c|c|c|}
\hline
$n_f$&$N_B$ &$N_{EJ}$&$N_{GLS}$\\\hline
0&-0.523$\pm$ 0.154&-0.523$\pm$0.154&-0.523$\pm$0.154\\ \hline
1&-0.487$\pm$ 0.126&-0.423$\pm$0.120&-0.479$\pm$0.121\\ \hline
2&-0.451$\pm$ 0.101&-0.291$\pm$0.085&-0.436$\pm$0.094\\ \hline
3&-0.414$\pm$ 0.079&-0.103$\pm$0.035&-0.393$\pm$0.070\\ \hline
4&-0.378$\pm$ 0.058&*&-0.351$\pm$0.059\\ \hline
5&-0.343$\pm$ 0.102&*&-0.311$\pm$0.134\\ \hline
6&-0.311$\pm$ 0.194&*&-0.272$\pm$0.232\\ \hline
\end{tabular}
\end{center}
\caption{\it Values of the infrared renormalon residue of the Bjorken, 
Ellis-Jaffe and 
GLS leading-twist Wilson 
coefficient $C_X$ after conformal mapping. "*" means that no stable 
result is obtained.}
\label{tableNX}
\end{table}

We are then able to give some estimates for the coefficients of the perturbative 
series. We provide them in tables~\ref{CB},~\ref{CEJS},~\ref{CGLS}. We should stress 
that our numbers incorporate the right asymptotic behaviour, which is not the case for 
large-$\beta_0$ estimates. For the Bjorken and GLS sum rules they have been calculated 
in \cite{Broadhurst:1993ru}.
The comparison with the exact result works reasonable well for $n_f$ smaller than 
6 for the Bjorken or GLS perturbative series. 
For large $n_f$ the comparison with the exact result gets worse. Note that 
for $n_f=6$, the normalization constant of the renormalon could be almost 
compatible with zero if the errors are included. This fits with the picture that 
the renormalon is less important when the number of flavours grows and one can 
reach to the point where the infrared renormalon disappears. For the Ellis-Jaffe 
perturbative series the same discussion applies with $n_f$ smaller than 4 (for $n_f=4$ 
no stable determination of the normalization constant can be obtained). Again this 
fit with the picture that the infrared renormalon becomes weaker when the number 
of active flavours increase.

We can also compare with other estimates one may find in the literature for the 
Bjorken (GLS) sum rules \cite{Kataev:1994gd,Kataev:1995vh,Ellis:1995jv,Contreras:2002kf}. 
We find that our predictions somewhat lie on the upper limit of the range of values 
obtained in these references for $C_{B/GLS}^{(3)}$. 

As a final comment it should be mentioned that the introduction of the leading logarithms in 
the renormalon, which is novel, does not actually improve the agreement with the running 
on $\nu$ of the perturbative coefficients of the perturbative series, when they are 
known. If this 
is an artifact of the leading-order analysis and would be solved at higher orders remains to be 
seen.  

\begin{table}[h!]
\hspace{-3cm}
\begin{center}
\begin{tabular}{|c|c|c|c|c|c|c|c|}
\hline
$n_f$&0 &1&2&3&4&5&6                                   \\\hline
-$C_{B}^{(0)}(Q)$& $0.523$ &  $0.487 $ & $0.451$ &$0.414$ & $0.378$ & $0.343$ & $0.311$ \\\hline
-$C_{B}^{(1)}(Q)$ &$ 0.516$ & $  0.452  $ & $  0.391 $ & $0.334 $ & $ 0.280  $ & $ 0.228 $ & $ 0.175$ \\\hline
-$C_{B}^{(2)}(Q)$ & $1.295 $ & $ 1.045$ & $ 0.824 $ & $0.630 $ & $ 0.461 $ & $  0.318  $ & $ 0.198  $ \\\hline
-$C_{B}^{(3)}(Q)$ & $4.199 $ & $ 3.149  $ & $ 2.285 $ & $1.588 $ & $ 1.042 $ & $ 0.630 $ & $0.334 $ \\\hline
-$C_{B}^{(4)}(Q)$ &$ 17.07 $ & $ 11.94 $ & $ 8.008 $ & $ 5.097 $ & $  3.023 $ & $  1.625 $ & $ 0.750 $ \\\hline
-$C_{B}^{(5)}(Q)$& $83.91 $ & $ 54.77 $ & $  34.08 $ & $ 19.93 $ & $  10.75 $ & $ 5.175 $ & $ 2.100$ \\ \hline
\end{tabular}
\end{center}
\caption{\it Renormalon-based estimates of the perturbative coefficients 
$C_{B}^{(s)}$ for $\nu=Q$ and for different number of flavours. We use the 
expression from Eq.~(\ref{CXRG1overn}) except for $C_{B}^{(0)}$ for which we use the expression
from Eq.~(\ref{CXRG}), otherwise the result is 0.}
\label{CB}
\end{table}
\begin{table}[h!]
\hspace{-3cm}
\begin{center}
\begin{tabular}{|c|c|c|c|}
\hline
$n_f$&1&2&3                                  \\\hline
-$C_{EJ}^{(0)}(Q)$& $ 0.423$ &$ 0.291$ & $ 0.103$\\ \hline
-$C_{EJ}^{(1)}(Q)$&  $ 0.392$ &$ 0.250$ & $ 0.080$\\ \hline
-$C_{EJ}^{(2)}(Q)$& $ 0.868$ & $ 0.478$ & $ 0.129$\\ \hline
-$C_{EJ}^{(3)}(Q)$&  $ 2.547$ & $ 1.253$ & $ 0.298$\\ \hline
-$C_{EJ}^{(4)}(Q)$&  $ 9.476$ & $ 4.222$ & $ 0.896$\\ \hline
-$C_{EJ}^{(5)}(Q)$& $ 42.87$ & $ 17.42$ & $ 3.336$ \\ \hline
\end{tabular}
\end{center}
\caption{\it Renormalon-based estimates of the perturbative coefficients 
$C_{EJ}^{(s)}$ for $\nu=Q$ and for different number of flavours. We do not display 
the column with $n_f=0$ since the numbers are equal to the Bjorken case.
We use the 
expression from Eq.~(\ref{CXRG1overn}) except for $C_{EJ}^{(0)}$ for which we use the expression
from Eq.~(\ref{CXRG}), otherwise the result is 0.}
\label{CEJS}
\end{table}
\begin{table}[h!]
\hspace{-3cm}
\begin{center}
\begin{tabular}{|c|c|c|c|c|c|c|}
\hline
$n_f$&1&2&3&4&5&6                                   \\\hline
-$C_{GLS}^{(0)}(Q)$&$ 0.479$ & $ 0.436$ & $0.393$ & $ 0.351$ & $ 0.311$ & $ 0.272$\\ \hline
-$C_{GLS}^{(1)}(Q)$& $ 0.449$ & $ 0.379$ & $ 0.317$ & $ 0.260$ & $ 0.206$ & $ 0.154$\\
\hline
-$C_{GLS}^{(2)}(Q)$& $ 1.029$ & $ 0.797$ & $ 0.598$ & $ 0.428$ & $ 0.288$ & $ 0.174$\\
\hline
-$C_{GLS}^{(3)}(Q)$& $ 3.100$ & $ 2.211$ & $ 1.507$ & $ 0.967$ & $ 0.569$ & $ 0.293$\\
\hline
-$C_{GLS}^{(4)}(Q)$& $ 11.75$ & $ 7.750$ & $ 4.838$ & $ 2.807$ & $ 1.469$ & $ 0.658$\\
\hline
-$C_{GLS}^{(5)}(Q)$& $ 53.92$ & $ 32.97$ & $18.92$ & $ 9.980$ & $ 4.680$ & $ 1.841$ \\ \hline
\end{tabular}
\end{center}
\caption{\it Renormalon-based estimates of the perturbative coefficients 
$C_{GLS}^{(s)}$ for $\nu=Q$ and for different number of flavours. We do not display 
the column with $n_f=0$ since the numbers are equal to the Bjorken case.
We use the 
expression from Eq.~(\ref{CXRG1overn}) except for $C_{GLS}^{(0)}$ for which we use the expression
from Eq.~(\ref{CXRG}), otherwise the result is 0.}
\label{CGLS}
\end{table}
\section{Renormalon subtracted scheme}
\label{RSsec}
In the previous section 
we have obtained the contribution to the perturbative series of the 
leading-twist Wilson coefficient that produces its leading asymptotic behaviour. This behaviour 
limits the accuracy that can be obtained for the observable 
from the perturbative series alone, which it will always have an error of the 
order of the higher twist corrections\footnote{Actually, this is so for the order 
in $\als$ for which the difference between the exact and the finite order 
result is 
minimal. If one goes to higher order in perturbation theory the series will deteriorate 
and the error will increase.}. Therefore, one can not obtain these 
higher-twist corrections unless the perturbative series is defined with power-like 
accuracy. The specific value for the higher twist will depend upon the specific prescription 
used. Here we will adapt the procedure used in Refs.~\cite{Pineda:2001zq,Pineda:2002se,Bali:2003jq}. 
In those references the non-analytic behaviour in the 
Borel plane that produced the asymptotic behaviour of the perturbative series was 
subtracted from the perturbative series and added to the subleading non-perturbative 
contributions. Actually the definition of the non-analytic piece is ambiguous and 
analytic terms can always be added.  The specific quantity we will substract to the 
perturbative series will be Eq.~(\ref{deltaCXRS}) with $n^*=1$\footnote{This is 
equivalent to what was called the RS' scheme in Ref.~\cite{Pineda:2001zq}, whereas the 
case $n^*=0$ was named the RS scheme. In our case here both schemes coincide, 
since the $n=0$ contribution from Eq.~(\ref{deltaCXRS}) vanishes because we expand the 
$\Gamma$'s in $1/n$. 
Obviously, we could also set $n^*$ different from 1 (as far as it is not too large). 
This would be equivalent to a change of scheme. The values of the non-perturbative 
matrix elements would change accordingly.}. This quantity fulfils the requirement that 
its imaginary part cancels the imaginary part of the perturbative series and that its 
dependence on $Q$ complies with the structure of the higher twist 
contribution. For illustration, the Bjorken sum rule would read
\be
\label{BjorkenRS}
M_1^p(Q^2)-M_1^n(Q^2)={g_A \over 6}C_{B,\RS}(Q;\nu_f)
-{4 \over 27}{1 \over Q^2}f_{3,\RS}(\nu_f)
\left[{\als(\nu_f^2)\over \als(Q^2)}\right]^{-\gamma_{\rm NS}^0 \over 2\beta_0}
\left(1+{\cal O}(\als)\right)
+{\cal O}\left({1 \over Q^4}\right)
\,,
\ee
where 
\be
\label{CBRS}
C_{B,\RS}(Q;\nu_f)=C_{B}(Q)-\delta C_{B,\RS}(\nu_f)=1+\sum_{s=0}^{\infty}C_{B,\RS}^{(s)}(Q/\nu;\nu_f/\nu)\als^{s+1}(\nu)
\,,
\ee
and
\be
\label{fRSf}
f_{3,\RS}(\nu_f)=f_{3}(\nu_f)
-{9\over 8}Q^2g_A
\left[{\als(Q)\over \als(\nu_f)}\right]^{-\gamma_{\rm NS}^0 \over 2\beta_0}
\delta C_{B,\RS}(Q;\nu_f)
\,.
\ee
Similar changes would apply to the Ellis-Jaffe and GLS sum rules. 
In Eq. (\ref{CBRS}) and Eq. (\ref{fRSf}), we have subtracted and added the contributions 
coming from the first infrared renormalon at the scale $\nu_f$ respectively. Thus, 
in this scheme, the  Wilson coefficient $C_{B,\RS}(Q;\nu_f)$ is free of the 
first infrared renormalon and the associated $n!$ behaviour.  
Therefore, the series is expected to converge better.
On the other hand, the higher twist, $f_{3,RS}(\nu_f)$, is free of the
first ultraviolet renormalon.  

In principle, the above series could be improved by incorporating the running 
in $\nu_f$ to any order in $\als$, which is 
renormalon free and therefore it could be obtained with good accuracy. 
In this situation it is legitimate to use the principal value (PV) prescription 
(or any other), since the renormalon ambiguity will cancel in the ratio. 
In this situation $\delta C_{X,\RS}$ reads
\be
\delta C_{X,\RS}^{(PV)}(Q;\nu_f)
=
\left(\als(Q) \over \als(\nu_f)\right)^{-b_X}
{\nu_f^2 \over Q^2}N_X \als(\nu_f)
\left[
D_{b+b_X}\left(-{4\pi \over \beta_0\als(\nu_f)}\right)
-1
\right]
\ee
where
\be
\label{DPV}
D_{b}(-x)=-xe^{-x}
\left\{x^b\cos(\pi b)\Gamma(-b)-(-x)^b\left[\Gamma(-b)-\Gamma(-b,-x)\right]
\right\}
\,,
\ee
and,
\be
\Gamma(b,x)=\int_x^\infty\! dt\, t^{b - 1} e^{-t}\,,
\ee
denotes the incomplete $\Gamma$ function. One would then work with 
the following quantity
\be
C_{B,\RS}(Q;\nu_f)=C_{B,\RS}(Q;\nu)
+
\left[\delta C_{X,\RS}^{(PV)}(Q;\nu)- \delta C_{X,\RS}^{(PV)}(Q;\nu_f)\right]
\,,
\ee
where one would set $\nu \sim Q$ and work order by order in perturbation theory 
in the first term in the right-hand side (where there is no large logs), whereas 
for the other terms the PV prescription is used. This allows to partially resum 
the dependence on $\nu_f$. Let us also note that 
the first term in Eq.~(\ref{DPV}) corresponds to $\Lambda_{\MS}$ (up to 
the anomalous dimension). Therefore it cancels in the difference,
$\delta C_{X,\RS}^{(PV)}(Q;\nu)- \delta C_{X,\RS}^{(PV)}(Q;\nu_f)$,
and can be neglected. Working with the principal value prescription 
has produced very good results in heavy quark physics, where it was 
possible to check the running of $\nu_f$ with "experiment" (lattice), see~\cite{Bali:2003jq}. However, 
in that situation, the running in $\nu_f$ was known with a very high accuracy, since there 
was no anomalous dimension and the running on $\nu_f$ could be deduced simply from the 
running of $\als$. In our case we only know the leading order and we expect a less 
accurate result.

Finally, we would like to mention that our knowledge of 
the running in $\nu$ of the higher twist terms is much more limited than 
in heavy quark physics. Here, we only know the leading log running. 
This has consequences in Eq.~(\ref{CBRS}) since, 
once we expand in $\als(\nu)$, there are some subleading logs which are unknown.

\section{Comparison with experiment}
In this section we will compare our theoretical predictions with the
experimental
data. Our aim is to perform a
combined global fit for the three sum rules. We
will use the experimental data for the Bjorken sum rule from
\cite{Abe:1998wq,Adeva:1998vw,Anthony:2000fn,Airapetian:2002wd,Deur:2004ti}
analyzed
according to Ref.~\cite{Deur:2004ti}. Note
that the elastic contribution has to be included in the experimental
numbers
in order the sum rule to be fully inclusive, i.e.
\be
\Gamma(Q^2)={1 \over
2}F_1(Q^2)\left(F_1(Q^2)+F_2(Q^2)\right)+\Gamma_{\rm inel.}(Q^2)
\,.
\ee
The empirical parameterization of the elastic form factors was taken
from~\cite{Mergell:1995bf}.
For large momentum this contribution is completely negligible.

We take the experimental data points for $M_1^p(Q^2)$ from Ref.~\cite{Osipenko:2005nx},
which have used the experimental results for the structure functions 
from~\cite{SLAC-E80,SLAC-E130,SLAC-E143,SLAC-E155,EMC-NA2,SMC-NA47,HERMES,CLAS}.

For the GLS sum rule we will use the experimental data from
the CCFR collaboration~\cite{Kim:1998ki}.

\begin{figure}[h!]
\begin{tabular}{cc}
\hspace{-5pt}
\put(-22,120){$M^B_1$}
\put(200,-6){$Q$}
\subfigure[Bjorken Sum Rule.]{\includegraphics[scale=0.98]{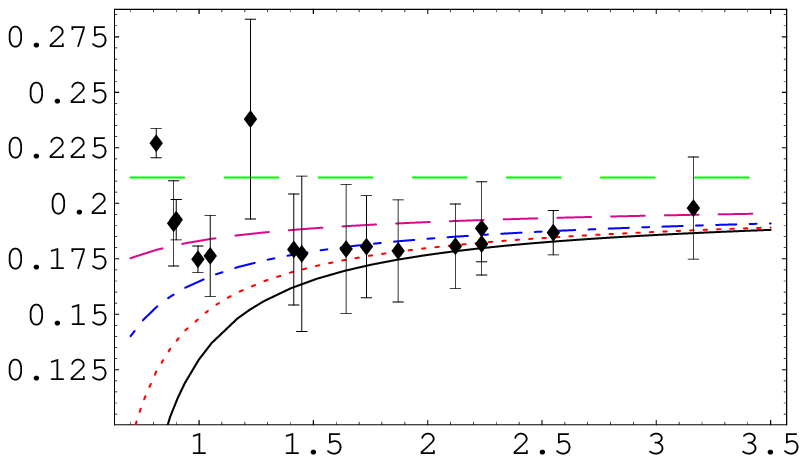}}&~~~
\put(-13,120){$M^p_1$}
\put(200,-6){$Q$}
\subfigure[Ellis-Jaffe Sum Rule.]{\includegraphics[scale=0.98]{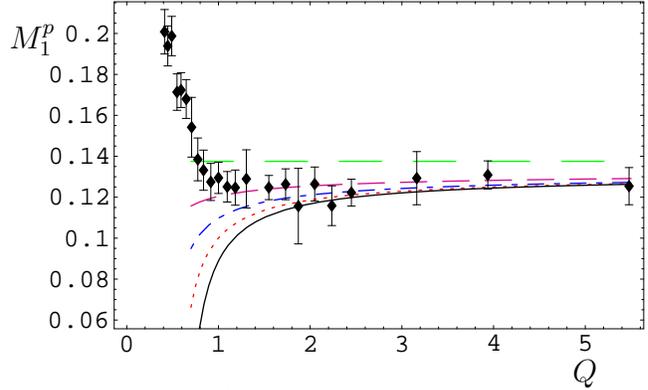}}
\end{tabular}
\begin{minipage}[t]{0.45\linewidth}
\hspace{10pt}
\begin{tabular}{c}
~\\
\put(-25,120){$M^{GLS}_3$}
\put(200,-6){$Q$}
\subfigure[GLS Sum Rule.]{\includegraphics[scale=0.90]{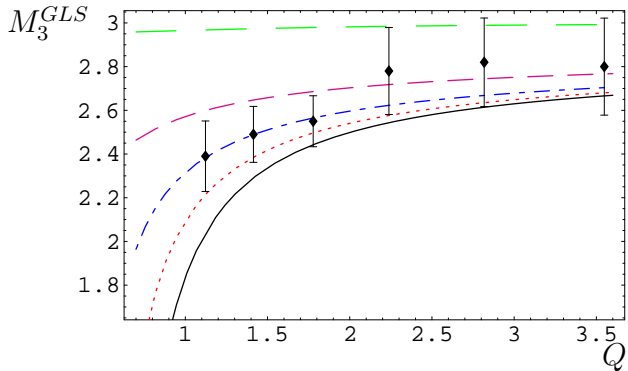}}
\end{tabular}
\end{minipage}
\hspace{0.6cm}
\begin{minipage}[t]{0.48\linewidth}
\vspace{-2cm}
\vspace{-17pt}
\caption{{\it Leading-twist contribution to the sum rules at different orders in perturbation theory in the $\MS$ scheme 
with $\nu=Q$ compared with the experimental data.
The long-dashed (green) line is the LO result. The dashed (magenta) line is the NLO result. The dot-dashed (blue) line 
is the NNLO result. The dotted (red) line is the NNNLO result and the continuous black line is the (estimate) N$^4$LO$^*$. At this 
order we have used a renormalon-based estimate for $C_X^{(3)}(Q)$ from Tables \ref{CB}, \ref{CEJS}, \ref{CGLS}. ${\hat a}_0=0.141$.}}
\label{figMS}
\end{minipage}
\end{figure}

\begin{figure}[ht!]
\hspace{-1pt}
\begin{tabular}{lr}
\put(200,-6){$Q$}
\put(-22,120){$M^B_1$}
\subfigure[{$\nu_f=0.8$.}]{\includegraphics[scale=0.95]{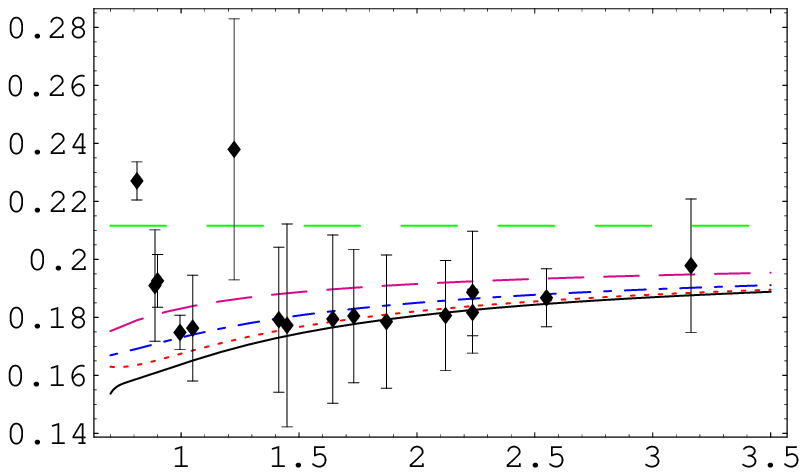}}
&~~~~~~
\put(-22,120){$M^B_1$}
\put(200,-6){$Q$}
\subfigure[{$\nu_f=1$.}]{\includegraphics[scale=0.95]{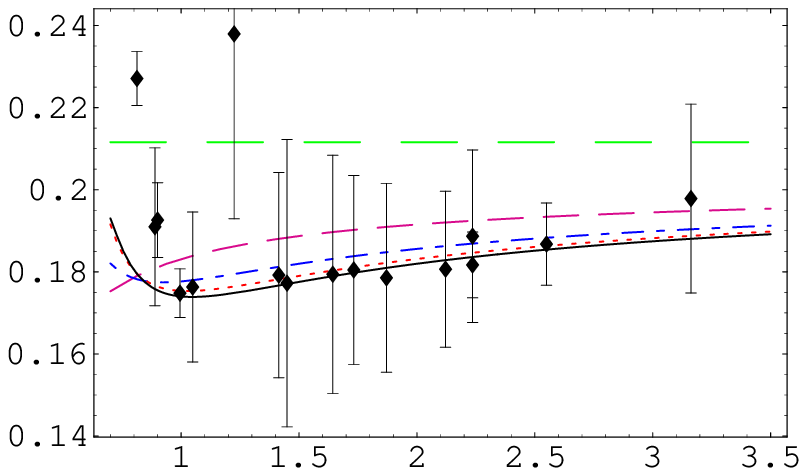}}
\vspace{-10pt}
\\
\put(-22,120){$M^{p}_1$}
\put(200,-6){$Q$}
\subfigure[{$\nu_f=0.8$.}]{\includegraphics[scale=0.95]{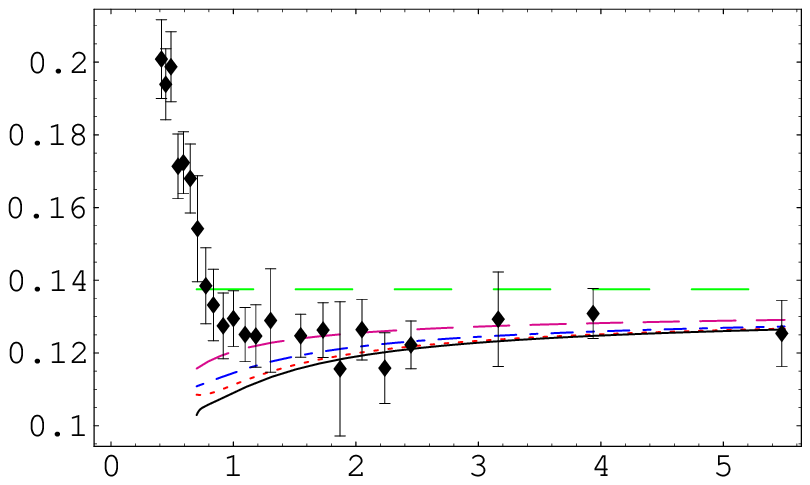}}
&~~~~~~
\put(-25,120){$M^{p}_1$}
\put(200,-6){$Q$}
\subfigure[{$\nu_f=1$.}]{\includegraphics[scale=0.95]{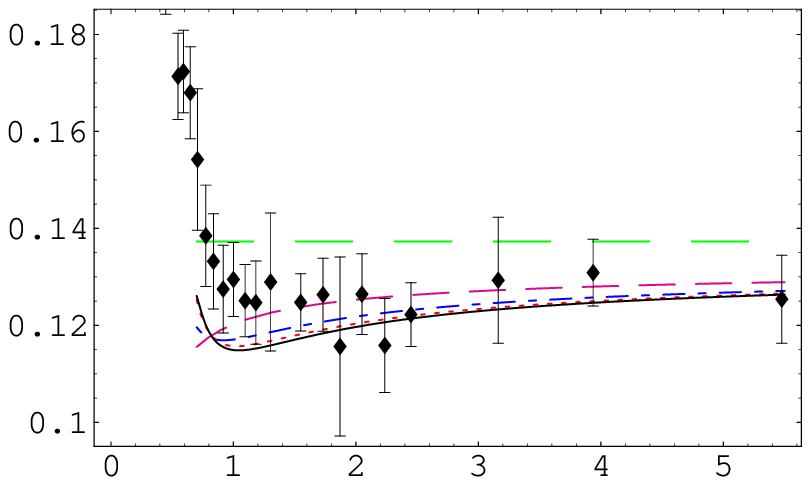}}
\vspace{-10pt}
\\
\put(-22,120){$M^{GLS}_3$}
\put(200,-6){$Q$}
\subfigure[{$\nu_f=0.8$.}]{\includegraphics[scale=0.95]{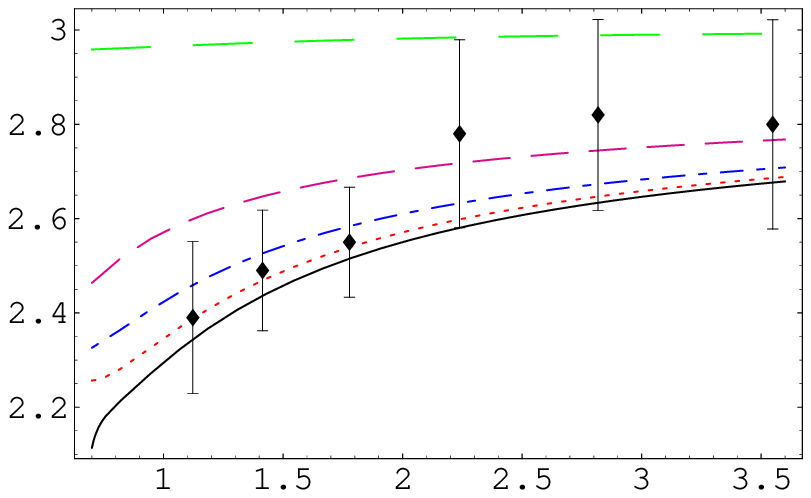}}
&~~~~~~
\put(200,-6){$Q$}
\put(-22,120){$M^{GLS}_3$}
\subfigure[{$\nu_f=1$.}]{\includegraphics[scale=0.95]{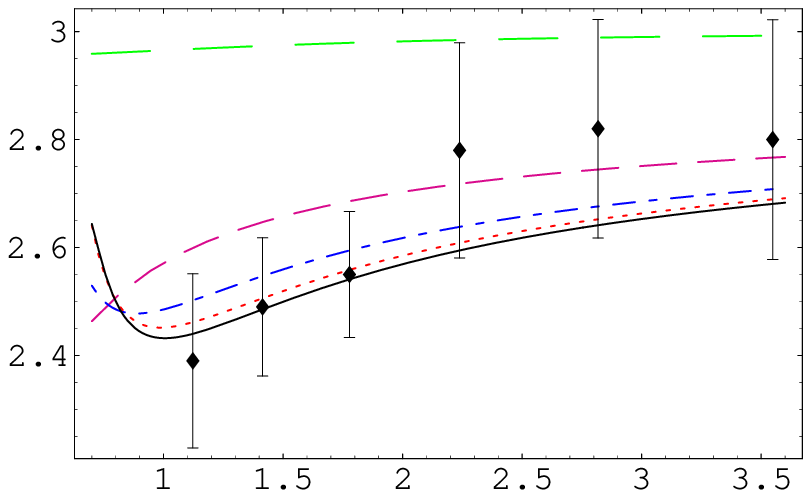}}
\end{tabular}
\caption{{\it 
Leading-twist contribution to the sum rules at different orders in perturbation theory in the $\RS$ scheme 
with $\nu=Q$ for two different values of the subtraction point, $\nu_f$, compared with the experimental data.
The long-dashed (green) line is the LO result. The dashed (magenta) line is the NLO result. The dot-dashed (blue) line 
is the NNLO result. The dotted (red) line is the NNNLO result and the continuous black line is the (estimate) N$^4$LO$^*$. At this 
order we have used a renormalon-based estimate for $C_X^{(3)}(Q)$ from Tables \ref{CB}, \ref{CEJS}, \ref{CGLS}. ${\hat a}_0=0.141$.}}
\label{RSGLS}
\end{figure}

We first illustrate the problem of convergence of the perturbative
series by
drawing the perturbative series in the $\MS$ scheme at different orders in
$\als$ in Fig.~\ref{figMS}.
The perturbative series has a relative good
convergence.
However, this convergence deteriorates when we approach to low energies.
We also see how the
perturbative theoretical result diverges from the experimental numbers
at low energies.
As we have already stated, the solution to this problem comes from using
the RS scheme. We plot again the perturbative series in the RS scheme in
Fig.~\ref{RSGLS} for two values of $\nu_f$: $\nu_f=0.8$ and 1 GeV.
We can see
that in both cases the convergence of the perturbative series improves.
This is specially so around the 1 GeV region. For $\nu_f=1$ GeV
we can also see a qualitative change in the figure with a much better
agreement with experiment around the 1 GeV region. This alone does not
mean much, since it only reflects the $\nu_f$ scale dependence of the
pure perturbative piece. Nevertheless, this scale dependence is known and
can be predicted by perturbation theory. This scale dependence
cancels with the scale dependence of the higher-twist terms. Therefore, the
complete result, including the higher twist terms, should be independent of
$\nu_f$. Thus, consistency demands that if we perform the fit to
experiment
including the higher twist $f_i(\nu_f)$ for different values of $\nu_f$,
in particular
for 0.8 and 1 GeV, the results obtained for the $f_i(\nu_f)$ should be
consistent with
the result obtained by performing the perturbative running in $\nu_f$
with respect
these two values. We have checked that this is so within the errors of
our evaluation.
This reassures the reliability of our fit. This also tells us that it is
reasonable to
use the operator product expansion formulas for $Q$ larger than 0.8 GeV.
This is the attitude
we will take in this paper where we will include (unless otherwise
indicated) the
experimental points for $Q >0.8$ GeV. Therefore, the perturbative plots
in Fig.~\ref{RSGLS}
with $\nu_f=1$ GeV can be reinterpreted as that a piece of the higher
twist correction
has been included in the pure perturbative term. This piece corresponds
to the
running of $\nu_f$ from 0.8 to 1 GeV of the higher twist term and can be
obtained from ´
the renormalization group. Therefore, in this sense, the change of slope
observed in
Fig.~\ref{RSGLS} can be interpreted as having its origin in perturbation
theory. Note that
the change of slope, from the experimental point of view, comes from the
elastic term.

In Fig.~\ref{RSGLS} higher-twist effects have not been included.
Our next aim is to perform a fit of ${\hat a}_0$ and the subleading twist
matrix elements from
the available experimental data. We refrain from trying to fit $\als$,
since
the experimental errors appear to be too large. Actually, they will be
one of the major
source of uncertainty of our analysis. We perform a global fit of all the
available data
from the different sum rules at the same time. The size of the
experimental errors
is the largest for the GLS sum rule, whereas the most accurate data come
from the
$M_1^p(Q^2)$ experimental points. In any case, $f_5^S$ will be obtained from
the GLS sum rule alone, since this fit is independent of the other two sum
rules. We perform the global fit to different orders in the expansion in
$\als$
of the leading twist perturbative Wilson coefficient. We work with the
running $\als$ consistent with the accuracy one is working at each 
order\footnote{We have also performed the analysis using the
four-loop running
$\als$ at any order. The final results are very similar. The use
of the
four-loop running $\als$ somewhat accelerates/improves the convergence
of the series.
Nevertheless, we prefer to keep ourselves consistent and only resum the
logs
associated to each order in perturbation theory. The fit using the
principal
value prescription is consistent with the finite order computation.
Nevertheless,
it is less precise because of the reasons mentioned in sec.~\ref{RSsec}.
Therefore,
we will not consider it further in this analysis.}. We show the results
of the
fit at different orders in perturbation theory in Table~\ref{Convergence}. We can
see how the series
shows convergence. We can then obtain relatively good estimates for
${\hat a}_0$ and $f_3$,
for the other non-perturbative parameters the situation is less
conclusive.
The values have been obtained with $\nu_f=1$ GeV.
\begin{table}[h!]
\hspace{-3cm}
\begin{center}
\begin{tabular}{|c|c|c|c|c|c|}
\hline
&${\hat a}_0$&$f_{0,\RS}$ (1 GeV)&$f_{3,\RS}$ (1 GeV)&$f_{5,\RS}^S$ (1 GeV)&$f_{8,\RS}$ (1 GeV)                                   \\\hline
LO&$0.043 $ & $ -1.571$ & $0.119$ & $1.030 $ & $5.90 $ \\ \hline
NLO& $ 0.120$ & $-0.817 $ & $ -0.075$ & $0.304 $ & $ 3.11$ \\
\hline
NNLO& $0.136 $ & $ 0.014$ & $ -0.106$ & $0.097$ & $ -0.18$ \\
\hline
NNNLO& $ 0.139 $ & $0.552$ & $-0.117 $ & $0.014$ & $-2.34$ \\
\hline
N$^4$LO$^*$& $ 0.141$ & $0.790 $ & $-0.124$ & $-0.029 $ & $ -3.30$ \\
\hline
\end{tabular}
\end{center}
\caption{\it Determination of ${\hat a}_0$ and the higher-twist non-perturbative parameters from the 
global fit to the sum rules at different orders in perturbation theory.}
\label{Convergence}
\end{table}

In order to estimate the errors, we allow for a variation of
$\als(M_z)=0.118\pm 0.003$,
of $N_X$ (according to the error given in Table~\ref{tableNX}), of the allowed set
of experimental points (we consider two situations: a) all the data
points for
$q > 0.8$ GeV and b) all the data points for $q > 1$ GeV; our central
values
will be the ones obtained with option a)), and also consider the
experimental errors.
We also consider the difference between the N$^3$LO$^*$ and NNLO result
as an estimate
of the error due to the convergence of the series. This error happens to be
very small in comparison with the other source of errors.
Summarizing, we obtain (with the $f_i$ in units of GeV$^2$)
\bea
{\hat a}_0&=&0.141^{+0.006}_{-0.004} (\delta \als) {}^{+0.002}_{-0.002}
(\delta N_X) {}^{+0.088}_{-0.088} ({\exp}) {}^{+0.001}_{-0.001} ({\rm pert})
 {}^{+0.010}_{-0.010}(q>1)
\,,
\\
f_{0,\RS}(1\;{\rm GeV})&=&0.790^{+0.241}_{-0.399} (\delta \als)
{}^{-0.489}_{+0.489} (\delta N_X) {}^{+0.159}_{-0.159} ({\exp}) 
{}^{+0.238}_{-0.238}({\rm pert}) {}^{+1.060}_{-1.060}(q>1)
\,,
\\
f_{3,\RS}(1\;{\rm GeV})&=&-0.124^{-0.050}_{+0.032} (\delta \als)
{}^{-0.049}_{+0.049} (\delta N_X) {}^{-0.121}_{+0.121} ({\exp}) 
{}^{+0.007}_{-0.007}({\rm pert}) {}^{+0.026}_{-0.026}(q>1)
\,,
\\
f_{5,\RS}^S(1\;{\rm GeV})&=&-0.029^{-0.162}_{+0.124}  (\delta \als)
{}^{-0.065}_{+0.065} (\delta N_X) {}^{-0.344}_{+0.344} ({\exp}) 
{}^{+0.042}_{-0.042} ({\rm pert}) \pm 0(q>1)
\,,
\\
f_{8,\RS}(1\;{\rm GeV})&=&-3.30^{-0.98}_{+1.61} (\delta \als)
{}^{+1.95}_{-1.95} (\delta N_X) {}^{-0.27}_{+0.27} ({\exp}) {}^{+0.96}_{-0.96}
({\rm pert}) {}^{+4.27}_{-4.27} (q>1)
\,.
\eea
If we combine all the
errors in quadrature we obtain
\bea
{\hat a}_0&=&0.141\pm 0.089
\,,
\label{a0final}
\\
f_{0,\RS}(1\;{\rm GeV})&=&0.790^{+1.225}_{-1.266}\; {\rm GeV}^2\,,
\label{f0final}
\\
f_{3,\RS}(1\;{\rm GeV})&=&-0.124^{+0.137}_{-0.142}\; {\rm GeV}^2\,,
\label{f3final}
\\
f_{5,\RS}^S(1\;{\rm GeV})&=&-0.029^{+0.374}_{-0.388}\; {\rm GeV}^2\,,
\label{f5final}
\\
f_{8,\RS}(1\;{\rm GeV})&=&-3.30^{+5.06}_{-4.90}\; {\rm GeV}^2\,,
\label{f8final}
\eea

We have also performed the fit with $\nu_f=0.8$. We obtain in this case
\bea
{\hat a}_0&=&0.143\,,
\\
f_{0,\RS}(0.8\;{\rm GeV})&=&0.365\,,
\label{f008}
\\
f_{3,\RS}(0.8\;{\rm GeV})&=&-0.210\,,
\label{f308}
\\
f_{5,\RS}^S(0.8\;{\rm GeV})&=&-0.167\,,
\label{f508}
\\
f_{8,\RS}(0.8\;{\rm GeV})&=&-1.58\,,
\label{f808}
\eea
whereas the magnitude of the errors is similar to the fit with $\nu_f=1$ GeV. 
As expected, the value of ${\hat a}_0$ almost remain independent of $\nu_f$.
The higher twist parameters do depend on $\nu_f$ in a way predicted 
by the renormalization group. For instance, for $f_3$ we would have the 
following expression\footnote{$\delta C_{B,\RS}$ has a renormalon 
ambiguity, which cancels in the difference we find in the 
second term in the right-hand side of the equation. We would like to remind 
the reader that in order to enforce this renormalon cancellation order 
by order in $\als$ both terms have to be expanded with $\als$ taken 
at the very same scale.}
\be
f_{3,\RS}(\nu_f)=f_{3,\RS}(\nu_f^{\prime})
\left[{\als(\nu_f^{\prime})\over \als(\nu_f)}\right]^{-\gamma_{\rm NS}^0 \over 2\beta_0}
+{9\over 8}\nu_f^2g_A
\left[\delta C_{B,\RS}(Q=\nu_f;\nu_f^{\prime})
- \delta C_{B,\RS}(Q=\nu_f;\nu_f)\right]
\ee
and analogously for the other higher twist terms. Therefore, we should recover 
the values in Eqs.~(\ref{a0final}-\ref{f8final}) after performing
the renormalization group running of Eqs.~(\ref{f008}-\ref{f808}). 
If we perform such running we obtain
\bea
f_{0,\RS}(1\;{\rm GeV})&=&0.335\,{\rm GeV}^{2},
\\
f_{3,\RS}(1\;{\rm GeV})&=&-0.130\,{\rm GeV}^{2},
\\
f_{5,\RS}^S(1\;{\rm GeV})&=&-0.004\,{\rm GeV}^{2},
\\
f_{8,\RS}(1\;{\rm GeV})&=&-1.46\,{\rm GeV}^{2}\,.
\eea
We see that the running goes in the right direction for $f_3$
and $f_5^S$, whereas for $f_0$ and $f_8$ the values remain constant. Either way, 
the numbers agree with those obtained in Eqs. (\ref{f0final}-\ref{f8final}) within errors.
\begin{figure}[h!!]
\begin{tabular}{l}
\put(315,-6){$Q$}
\put(77,120){$M^B_1$}
\hspace{3.5cm}\subfigure[Bjorken Sum Rule.]{\includegraphics[scale=1]{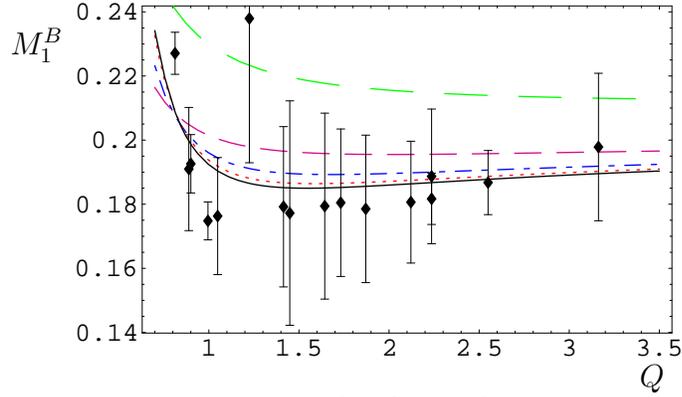}}
\vspace{-9pt}
\\
\put(315,-6){$Q$}
\put(77,120){$M^{p}_1$}
\hspace{3.5cm}
\subfigure[Ellis-Jaffe Sum Rule.]{\includegraphics[scale=1]{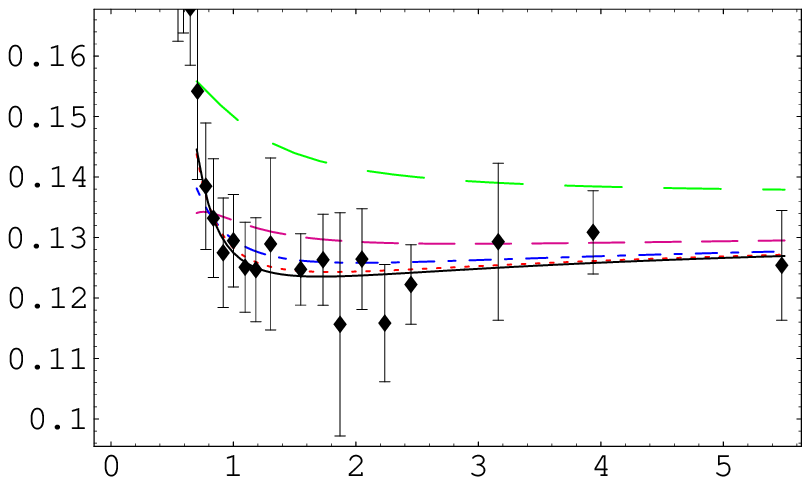}}
\vspace{-9pt}
\\
\put(315,-6){$Q$}
\put(77,120){$M^{GLS}_3$}
\hspace{3.5cm}
\subfigure[GLS Sum Rule.]{\includegraphics[scale=1]{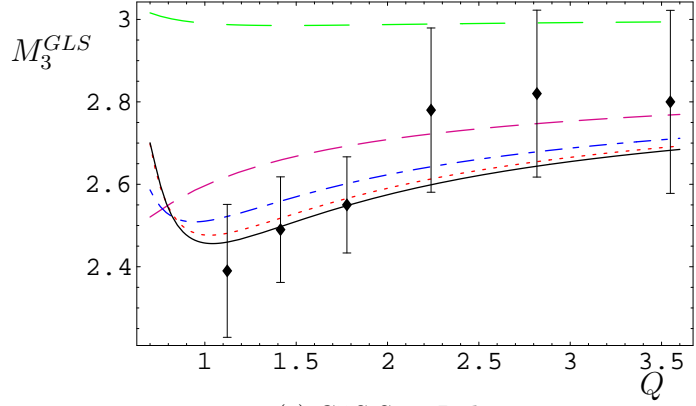}}
\end{tabular}
\caption{{\it Global fit to the Sum Rules in the RS scheme with $\nu=Q$ at different orders in perturbation theory 
including the leading and next-to-leading twist term. The long-dashed (green) line is the LO result. The dashed (magenta) 
line is the NLO result. The dot-dashed (blue) line is the NNLO result. The dotted (red) line 
is the NNNLO result and the continuous black line is the (estimate) N$^4$LO$^*$. At this 
order we have used a renormalon-based estimate for $C_X^{(3)}(Q)$ from Tables \ref{CB}, \ref{CEJS}, \ref{CGLS}. 
The values of ${\hat a}_0$, $f_i$ and $f_5^S$ are taken from Eqs.  (\ref{a0final}-\ref{f8final}).}}
\label{FittotalGLS}
\end{figure}

Finally, we have also considered the inclusion of $1/Q^4$ corrections.
The values of $f_0$ and $f_8$ are not stable under
the inclusion of these effects, although there is a correlation on
the values of $f_0$ and $f_8$: a large positive value
of $f_0$ is only possible if we also have a large negative value of
$f_8$. The point is that, even if formally it should be
possible to distinguish $f_0$ and $f_8$ due to the different anomalous
dimension, in practice we do not have enough
accuracy. Therefore, the values obtained above for $f_0$ and $f_8$ (and
its errors) should be taken with caution. This warning also 
applies to the determination of $f_5^S$, for which the inclusion of 
$1/Q^4$ corrections significantly changes its value. 
For the other coefficients, ${\hat a}_0$ and $f_3$ the variations are smaller than the errors of
our fit. 

The error appears to be dominated by the experimental one in ${\hat a}_0$ and
$f_3$, the objects we can compute with better accuracy.

We would also like to note that, in some cases, the values of the higher-twist non-perturbative 
parameters are compatible with zero within errors. 

To illustrate the quality of the fit we plot our final results, 
the sum rules including the 
leading and subleading twist, at different
orders in perturbation theory with our best fit, Eqs. 
(\ref{a0final}-\ref{f8final}), 
compared with the experimental data in Fig.~\ref{FittotalGLS}.

The analysis of the GLS sum rule taking into account renormalon
effects has been considered in Ref.~\cite{Contreras:2002kf}.
In this reference only two data points were used and a
Principal Value-like Borel resummation prescription was used.
The authors also included some pure non-perturbative effects
arguing that they could be inherited from the renormalon computation.
This is still an open question and, actually, it has been criticised
by one of the authors in Ref.~\cite{Cvetic:2002qf}.

\section{Conclusions}

We have studied the large order behaviour in perturbation theory of the
Bjorken, Ellis-Jaffe and GLS sum rules. In particular, 
we have considered their first infrared renormalons, for which we have 
obtained their analytic structure with logarithmic accuracy and also an 
approximate determination of their normalization constant. 
Estimates of higher order terms of the perturbative series are
given. The RS scheme has been worked out for these
observables and compared with experimental data. The convergence of the 
perturbative series
greatly improves in this scheme, specially around the 1 GeV region. In 
particular,
for $\nu_f=1$ GeV, the agreement between the pure perturbative contribution
and experiment is quite good. The fact that we have a convergent series in
perturbation theory allows to give meaningful values for the higher twist
condensates with well defined errors. We have performed a detailed 
analysis, being
able to give predictions for ${\hat a}_0$ and some higher twist condensates, 
including error
bars. Our best fits for the sum rules can be found in Fig.~\ref{FittotalGLS}. 
The values for the non-perturbative matrix elements read
\bea
{\hat a}_0&=&0.141\pm 0.089
\,,
\\
f_{3,\RS}(1\;{\rm GeV})&=&-0.124^{+0.137}_{-0.142}\,{\rm GeV}^{2}
\,.
\eea
The experimental situation is not very good for the GLS sum rule,
for which we can not give a precise number for the higher twist.
The experimental precision is not good enough to check the assumption 
of~\cite{Kataev:2005ci} that $f_3$ and
$f_5^S$ are equal. We also do not display here the values of $f_0$ and $f_8$,
since they have large errors and their values in the fit are somewhat
correlated. A large value of one of them could be obtained in the fit to
the price of having the other being large with opposite sign.

One of the most important source of error of the present analysis 
is the experimental one. Any 
improvement
in this respect will immediately lead to a reduction of the errors of 
the numbers
obtained in this paper.

The quality of the analysis is worse than the one obtained 
for heavy quark physics analysis. The 
sensitivity 
to the renormalon is smaller and the determination of the normalization 
constant
is less accurate that in that case. This was to be expected since 
the singularities in the Borel plane are more  far away here than in heavy 
quark physics. In any case, the inclusion of the 
renormalon
cancellation introduces a qualitative change on the perturbative 
behaviour around
the 1 GeV region making it much closer to the experimental figures.

Another issue we would like to mention is that the resummation of 
renormalon-related logarithms does not appear to improve the convergence of the 
series. On the other hand, we have only performed the leading log 
resummation in this paper. It would be interesting to see what happens 
at higher orders.  

Finally, the possibility to merge with the chiral limit seems closer now but the 
gap still
exists. In particular one should find a systematic way to incorporate 
higher twist effects.

\medskip
   
\noindent
{\bf Acknowledgments.}\\
 We would like to thank A. Deur and M. Osipenko 
for providing detailed information on the experimental points of Refs. 
\cite{Deur:2004ti} and~\cite{Osipenko:2005nx} respectively. 
F.C.\ acknowledges support of the Natural Sciences and Engineering
Research Council of Canada. A.P.\ is supported by MCyT and Feder (Spain),
FPA 2004-04582-C02-01 and by CIRIT (Catalonia), 2001SGR-00065.
%
%

\end{document}